\documentclass{mn2e}
\usepackage{psfig}

\newcommand{\etal}{et~al.}

\newcommand{\ionhy}{H{\sc ii}}

\newcommand{\water}{$\mbox{H}_{2}\mbox{O}$}

\newcommand{\kms}{$\mbox{km~s}^{-1}$}

\newcommand{\dfig}[2]            
{
  \begin{center}
    \begin{minipage}[t]{0.45\textwidth}
        \psfig{file=./#1.eps,height=0.75\textwidth,angle=270}
    \end{minipage}
    \hfill
    \begin{minipage}[t]{0.45\textwidth}
        \psfig{file=./#2.eps,height=0.75\textwidth,angle=270}
    \end{minipage}
  \end{center}
}

\newcommand{\sfig}[1]            
{
  \begin{center}
    \begin{minipage}[t]{0.45\textwidth}
        \psfig{file=./#1.eps,height=0.75\textwidth,angle=270}
    \end{minipage}
  \end{center}
}

\begin{document}

\title[Search for 4750- and 4765-MHz OH masers] {A search for 4750- and
  4765-MHz OH masers in Southern Star Forming Regions}

\author[Dodson \& Ellingsen]{R.G. Dodson $^1$,
  S.P. Ellingsen $^1$ \\
$^1$ School of Mathematics and Physics, University of Tasmania, 
     GPO Box 252-21, Hobart, Tasmania 7001, Australia;\\  
     Richard.Dodson@utas.edu.au, Simon.Ellingsen@utas.edu.au\\}

\maketitle

\begin{abstract}
  
  We have used the Australia Telescope Compact Array (ATCA) to make a
  sensitive (5-$\sigma$ $\simeq$ 100 mJy) search for maser emission from
  the 4765-MHz $^2\Pi_{1/2}$ F=1$\rightarrow$0 transition of OH.
  Fifty five star formation regions were searched and maser emission
  with a peak flux density in excess of 100~mJy was detected toward
  fourteen sites, with ten of these being new discoveries. In addition
  we observed the 4750-MHz $^2\Pi_{1/2}$ F=1$\rightarrow$1 transition
  towards a sample of star formation regions known to contain 1720-MHz
  OH masers, detecting marginal maser emission from G348.550-0.979. If
  confirmed this would be only the second maser discovered from this
  transition.
  
  The occurrence of 4765-MHz OH maser emission accompanying 1720-MHz
  OH masers in a small number of well studied star formation regions
  has lead to a general perception in the literature that the two
  transitions favour similar physical conditions. Our search has found
  that the presence of the excited-state 6035-MHz OH transition is a
  much better predictor of 4765-MHz OH maser emission from the same
  region than is 1720-MHz OH maser emission. Combining our results
  with those of previous high resolution observations of other OH
  transitions we have examined the published theoretical models of OH
  masers and find that none of them predict any conditions in which
  the 1665-, 6035- and 4765-MHz transitions are simultaneously
  inverted.

\end{abstract}


\section{Introduction}

High-mass star formation regions frequently exhibit OH maser emission,
with thost most commonly observed being from the main-lines of the
ground state $^2\Pi_{3/2} (J=3/2)$ transitions, in particular the
1665-MHz ($F=1\rightarrow 1$) transition. Masing in the other ground
state main-line transition at 1667~MHz ($F=2\rightarrow 2$) frequently
accompanies the 1665-MHz masers, and in some sources the 1612- or
1720-MHz satellite-line transitions also exhibit maser action.
Emission from the OH molecule has been detected towards star formation
regions for all levels with energies less than approximately 500~K
above the ground state.  A number of searches towards star-forming
regions have been undertaken for the higher exited OH transitions at
4765~MHz \cite{GM83,CMW91,CMC95,SKH00} and 6035~MHz
\cite{BDWC97,S94,CV95} (the $^2\Pi_{1/2}$, F=1$\rightarrow$0 and
$^2\Pi_{5/2}$, F=3$\rightarrow$3 transitions respectively).

For the $^2\Pi_{3/2} (J=3/2)$ and $(J=5/2)$ levels, the most commonly
observed masing transitions towards star-forming regions are in each
case one of the two main lines. However, for the $^2\Pi_{1/2} (J=1/2)$
level, which is the lowest energy level of the $^2\Pi_{1/2}$ ladder,
emission from the single main-line transition at 4750-MHz is very rare
and the most commonly observed transition from this level is the F=
$1\rightarrow0$ satellite line, analogous to the 1720-MHz transition
in the ground-state. Masing emission from all of the transitions in
the $^2\Pi_{1/2}$ ladder is rare and one obvious reason for this is
that all the levels are at a higher energy than the two lowest levels
in the $^2\Pi_{3/2}$ ladder level where masing emission is relatively
strong and common. The reason for the 4765-MHz transition being the
most common masing transition in the $^2\Pi_{1/2}$ ladder is that the
$F=1$ levels in $^2\Pi_{1/2} (J=1/2)$ naturally become over populated
with respect to the $F=0$ due to selection rules \cite{E77}. 4765-MHz
OH maser emission typically takes the form of a single narrow masing
line and they are generally considered to be more highly variable than
ground-state transitions, the most spectacular example being Mon~R2
\cite{SCH98}, which had a peak flux of 80~Jy in late 1997, but which
we were unable to detect (5-$\sigma$ = 80~mJy) in these observations.
Intriguingly, Mon~R2 is also unique in that it is the only 4765-MHz OH
maser which has ever been found to exhibit any polarization
\cite{SCH98}. Monitoring of a number of other 4765-MHz OH masers
\cite{S97,SKH00} shows that typically the degree of variability is
much less than that for Mon~R2. Analysis of the line width of the
masers during variation by Szymczak \etal\/ suggests that the masers
are saturated, as do VLBI observations by Baudry \etal\/
\shortcite{BDBGW88}.

Towards a small number of well studied OH maser sources (NGC7538,
Sgr~B2, W3(OH) \& Mon~R2) the 1720- and 4765-MHz OH maser transitions
have been observed to be spatially coincident to within the relative
positional errors of the observations
\cite{PGW84,GWP87,BDBGW88,GCRYF01,SCH98}. However, in most of these
cases the positional errors are relatively large and the emission at
the two frequencies is not necessarily coincident in velocity once
Zeeman splitting is taken into account. The apparent
association between 1720- and 4765-MHz OH masers was further
strengthened by the results of the search of Cohen \etal\/
\shortcite{CMC95}. They searched a sample of {\em IRAS} sources
bright at 60~$\mu$m and found that all but one of the sites
exhibitting 1720-MHz emission also showed 4765-MHz emission. MacLeod
\shortcite{M97} examined the relationship between 1720-, 4765- and
6035-MHz OH masers, again finding a strong correlation between the
presence of 4765- and 1720- MHz OH masers. His analysis also shows that the vast majority of 4765-MHz maser
sources exhibit 6035-MHz maser emission (there being only two
exceptions).

While radiative pumps are generally favoured for excitation of the
ground-state OH lines, it is possible to produce inversion in the
4765-MHz transition through either radiative or collisional processes
\cite{E77}. Current theories prefer radiative excitation of all OH
maser lines including those from the $^2\Pi_{1/2}$ ladder
\cite{PK00,GCRYF01}, although in the past collisional models have been
put forward \cite{KN90}. Much of the focus of previous modelling of
4765-MHz OH masers has been on finding conditions under which it is
inverted at the same time as the 1720-MHz transition. Combining our
observations of the $^2\Pi_{1/2}$, F=1$\rightarrow$1 and
$^2\Pi_{1/2}$, F=1$\rightarrow$0 transitions with those of the
ground-state and 6035-MHz transitions by Caswell
\shortcite{C97,C98,C99}, (also observed at high-resolution with the
ATCA), we have investigated the associations between the various OH
transitions. These observations constitute a unique set of data, as
for the first time a large number of sites have been observed in all
the common OH maser transitions, at arcsecond resolution, with good
sensitivity using the same instrument. Determining whether the
different OH maser transitions are coincident is difficult as it
requires full polarization VLBI observations to obtain the necessary
spatial resolution and to account for the velocity shifts due to
Zeeman splitting. From our observations and those of Caswell we are
not able to determine coincidence of individual maser spots, however,
they are able to show if the different transitions occur within the
same maser cluster and whether the velocities are aligned.

It was suggested more than 30 years ago that since there are a number
of OH maser transitions which are relatively common in star-formation
regions, it should be possible to use observations of multiple
transitions to test and constrain theoretical models of OH maser
emission \cite{ZP70}. The only source which has been studied at high
spatial resolution in a large number of OH transitions is W3(OH)
\cite{RHBMJS80,BDBGW88,MFGMCB94,BM95,BD98}, for which Cesaroni \&
Walmsley \shortcite{CW91} produced a model which qualitatively matches
the masers, thermal emission and absorption observed from the OH
molecule towards the source. A similar approach has recently been
used with some success for multiple methanol maser transitions towards
sites of high-mass star formation \cite{CSECGSD01,SSECMOG01} and the
close relationship between OH and methanol masers \cite{C97} holds out
the possibility of combining observations of both species to further
constrain pumping models and the implied physical conditions
\cite{CSG02}. Observations such as these can be compared with the
detailed studies of individual sources to determine to what degree the
conditions inferred are generally applicable.

\section{Observations and Data reduction} 

The Australia Telescope Compact Array was used to make the
observations of both the 4750- and 4765-MHz transitions, with rest
frequencies 4750.656~MHz and 4765.562 respectively \cite{DMBB77}. The
configuration of six antennas yielded 15 baselines between 5 and 95
k$\lambda$ for the 4765-MHz and between 1 and 90 k$\lambda$ for the
4750-MHz observations. All pointings had 30 minutes or more of
on-source observation, and the primary (and bandpass) calibrations
were obtained from observations of PKS~1934-638. The correlator was
configured with a 4-MHz bandwidth and 1024 spectral channels for each
of the four polarisation products recorded. The sources were selected
from those observed to exhibit either 6035- or 1720-MHz OH masers
\cite{C97,C99}, or those previously reported to have 4765-MHz emission
\cite{CMW91,CMC95,M97,S97}. The observations spanned two epochs,
25-26 August 1999 for the 4750-MHz observations and 15-16 September
2000 for the 4765-MHz observations. Only those sites which showed
1720-MHz or 4765-MHz emission were searched for 4750-MHz
emission. There are no reported southern sites of 4750-MHz emission
other than that in Sgr~B2 \cite{GWP87}.

Data reduction was performed using the {\bf miriad} and {\bf karma}
software packages and following the standard methods. With 1024
channels the effective velocity resolution at 4765-MHz is 0.3\kms.
The 1-$\sigma$ level over this bandwidth was $\approx$ 20~mJy per
channel and the positional accuracy of the observations was typically
0.6\arcsec. We used a number of different methods to search for maser
emission in our interferometric dataset. A vector averaged spectrum
with a velocity resolution of 0.3~\kms\/ was formed at the positions
listed in tables~\ref{tab:nondet4750}~\&~\ref{tab:nondet4765}. The
velocity range covered by the spectrum and the measured 5-$\sigma$
noise levels are listed in the same tables. In addition two image
cubes were produced (both centred on the velocity listed in the
tables), the first with an angular size of 256\arcsec\ covering a
velocity range of $\pm$5~\kms with a 0.2~kms\/ resolution and the
second cube with an angular size of 150\arcsec\ and a velocity range
of $\pm$20~\kms\/ with a resolution of 1~\kms. The 1-$\sigma$ noise
levels in each of these images were approximately 30 and 12~mJy/beam
respectively.

Unlike the ground-state OH transitions the $^2\Pi_{1/2} (J=1/2)$
transitions are not expected to exhibit circular polarisation, as this
level is diamagnetic (unlike all other OH rotational energy levels
which are paramagnetic). None of the 4765-MHz masers observed have
detectable levels of linear, or circular polarisation, the most
stringent upper limits being 3.6\% for linear polarisation and 5\% for
circular polarisation.

\section{Results}

\subsection{4750-MHz transition}
\begin{figure}
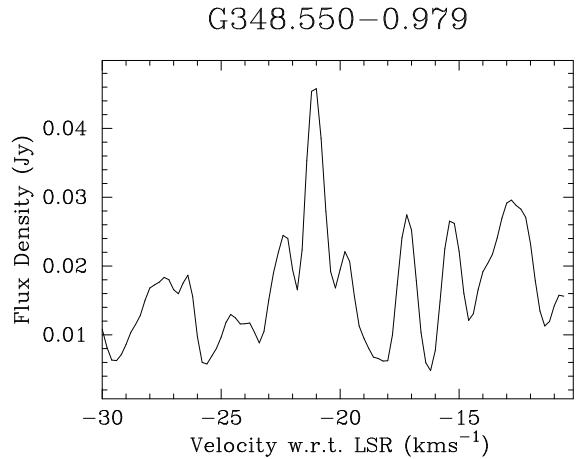

\sfig{g348.550-00.979}
\caption{Spectrum of the marginal 4750-MHz OH maser detection towards
  G348.550-0.979}
\label{fig:348_4750}
\end{figure}

A total of 17 star forming regions were searched for 4750-MHz OH
emission, the position, velocity range and noise level for each site
is given in Table~\ref{tab:nondet4750}.  Only one of the 12 sites with
1720-MHz OH masers showed any sign of 4750-MHz emission, this being
the marginal detection towards G348.550-0.979
(Fig.~\ref{fig:348_4750}). The possible weak 4750-MHz OH maser
emission at -21\kms\/ does not match the velocity of the 1720- or
6035-MHz masers towards this site \cite{C99,CV95}, but does overlap
that of the 1665- and 1667-MHz masers. The 4750-MHz transition has
received very limited attention, with the only major searches being
those by Gardner \& Martin-Pintado \shortcite{GM83} and Cohen \etal\/
\shortcite{CMW91,CMC95}. The only site which has previously been
reported to show 4750-MHz OH maser emission is S252 which was detected
as an OH maser by Cohen \etal\/ \shortcite{CMW91} and which subsequent
observations showed to be quite variable \cite{CMC95}.  One of the
sources searched for 4750-MHz emission was Sgr~B2, which was the first
reported source to show emission from this transition \cite{GR71}.
Interferometric observations of the same source with a synthesised
beamwidth of approximately 3\arcsec\/ show a similar spectrum
\cite{GWP83}. Our observations are at a higher spatial resolution and
a combination of this and a relatively short integration time have
produced a larger than average noise level for Sgr~B2 which has meant
that the 4750-MHz emission in this source is beneath our detection
limit.

\begin{table*}
  \caption{Sources searched for 475O-MHz OH maser emission. The 5-$\sigma$ 
    limit column is for emission in a vector averaged spectrum at centred on 
    the position, and covering the velocity range listed. References :
    a = Caswell \protect\shortcite{C97};
    b = Caswell \protect\shortcite{C99};
    c = Cohen \etal\/ \protect\shortcite{CMC95};
    d = MacLeod \protect\shortcite{M97};
    e = Smits \etal\/ \protect\shortcite{SCH98}}
  \begin{tabular}{lcccrcl}
    \hline
               & \multicolumn{2}{c}{\bf Reference position} & {\bf Velocity} & 
    {\bf 5-$\sigma$} & {\bf Central}  &
                     \\
  {\bf Source} & {\bf Right Ascension} & {\bf Declination} & {\bf Range}    &
    {\bf Limit}           & {\bf Velocity} &
                     \\
  {\bf Name}   & {\bf (J2000)}         & {\bf (J2000)}     & {\bf (\kms)}   &
    {\bf (Jy)}            & {\bf (\kms)}   &  
    {\bf References} \\ \hline
  Mon R2            & 06:07:47.832 & -06:22:56.43& -40--60 & 0.2  &-10.7       & e   \\
  G294.511$-$1.621& 11:35:32.208 & -63:14:43.00& -60--40 & 0.1 &-12,-12        & a,b \\
  {\em IRAS}12073$-$6233& 12:10:00.360 & -62:49:57.00& -20--80 & 0.4  &32      & c   \\
  G306.322$-$0.334& 13:21:22.950 & -63:00:28.40& -73--27 & 0.1 &-23            & b,d \\
  G338.925$+$0.557& 16:40:33.490 & -45:41:37.00& -112-- -12 & 0.1 &-61.3       & b,d \\
  G339.622$-$0.121& 16:46:06.030 & -45:36:43.90& -87--12 & 0.1 &-36.7          & b,d \\
  G340.785$-$0.096& 16:50:14.838 & -44:42:26.71& -146-- -52 & 0.1 &-105.6,-102 & a,b \\
  G345.003$-$0.224& 17:05:11.205 & -41:29:06.96& -75--24 & 0.1 &-26.2,-29.3    & a,b \\
  G348.550$-$0.979& 17:19:20.418 & -39:03:51.65& -63--36 & 0.1 &-13,-12.1      & a,b \\
  G351.417$+$0.645& 17:20:53.370 & -35:47:01.16& -60--39 & 0.2 &-10,-10.4      & a,b \\
  {\em IRAS}17175$-$3544& 17:20:54.945 & -35:47:02.38& -60--39 & 0.2 &-10      & c   \\
  G351.775$-$0.536& 17:26:42.559 & -36:09:15.99& -50--49 & 0.1 &-8.7,4         & a,b \\
  G353.410$-$0.360& 17:30:26.189 & -34:41:45.74& -70--29 & 0.2 &-20.7,-19      & a,b \\
  G0.666$-$0.029  & 17:47:18.640 & -28:22:54.60& 22--120 & 0.5 &72.6           & a   \\
  G0.666$-$0.035  & 17:47:20.143 & -28:23:06.18& 22--120 & 0.5 &67             & a   \\
  G10.473$+$0.027 & 18:08:38.330 & -19:51:47.90& 10--109 & 0.2 &64.7           & b   \\
  G11.034$+$0.062 & 18:09:39.770 & -19:21:23.60& -30--60 & 0.1 &22.5           & b   \\
\hline
  \end{tabular}
  \label{tab:nondet4750}
\end{table*}

\subsection{4765-MHz transition}
\begin{table*}
  \begin{minipage}{\textwidth}
  \caption{Sources searched for 4765-MHz OH maser emission. The 5-$\sigma$ 
    limit column is for emission in a vector averaged spectrum at centred on 
    the position, and covering the velocity range listed. References :
    a = Caswell \protect\shortcite{C97} ; 
    b = Caswell \protect\shortcite{C98} ;
    c = Caswell \protect\shortcite{C99} ;
    d = Cohen \etal\/ \protect\shortcite{CMC95} ; 
    e = Gardner \& Martin-Pintado \protect\shortcite{GM83};
    f = Smits \etal\/ \protect\shortcite{SCH98}.} 
  \begin{tabular}{lcccrcl}
    \hline
               & \multicolumn{2}{c}{\bf Reference position} & {\bf Velocity} & 
    {\bf 5-$\sigma$} & {\bf Central}  & 
                     \\
  {\bf Source} & {\bf Right Ascension} & {\bf Declination} & {\bf Range}    &
    {\bf Limit}           & {\bf Velocity} & 
                     \\
  {\bf Name}   & {\bf (J2000)}         & {\bf (J2000)}     & {\bf (\kms)}   &
    {\bf (Jy)}            & {\bf (\kms)}   & 
    {\bf References} \\ \hline
  Mon~R2        & 06:07:47.832 & -06:22:56.43& -78--163  & 0.08 & -10.7     & f  \\ 
  G240.316$+$0.071& 07:44:51.971 & -24:07:42.32& -90--150  & 0.08 &  63.6     & a  \\ %
  G284.351$-$0.418& 10:24:10.680 & -57:52:34.01& -103--138 & 0.11 &   4.7     & a  \\
  G285.263$-$0.050& 10:31:29.884 & -58:02:18.48& -103--138 & 0.09 &   9.3     & a  \\
  G294.511$-$1.621& 11:35:32.208 & -63:14:43.00& -105--135 & 0.09 & -12,-12   & a,c  \\ 
  {\em IRAS}12073-6233& 12:10:00.360 & -62:49:57.00& -106--135 & 0.09 &  32.0     & d  \\
  G300.969$+$1.148& 12:34:53.272 & -61:39:39.91& -106--134 & 0.09 & -37.8     & a  \\
  G305.200$+$0.019& 13:11:16.881 & -62:45:54.72& -107--134 & 0.09 & -33.2     & a  \\
  G306.322$-$0.334& 13:21:22.950 & -63:00:28.40& -107--133 & 0.09 & -23.0     & c  \\ 
  G309.921$+$0.479& 13:50:41.773 & -61:35:10.08& -107--133 & 0.08 & -59.8     & a  \\ %
  G311.596$-$0.398& 14:06:18.353 & -62:00:15.29& -108--133 & 0.10 &  31.4     & a  \\
  G311.643$-$0.380& 14:06:38.742 & -61:58:23.15& -108--132 & 0.09 &  33.8     & a  \\
  G323.459$-$0.079& 15:29:19.332 & -56:31:21.26& -108--132 & 0.09 & -70.2     & a  \\
  G328.307$+$00.43& 15:54:06.437 & -53:11:41.11& -108--132 & 0.11 & -90.4     & a  \\ %
  G328.808$+$0.633& 15:55:48.387 & -52:43:06.69& -108--132 & 0.10 & -45.7,-45,-43.5 &a,a,c \\ %
  G331.511$-$0.102& 16:12:09.756 & -51:28:37.84& -108--132 & 0.11 &  -90.0    & a  \\
  G331.512$-$0.103& 16:12:09.926 & -51:28:37.06& -108--132 & 0.12 &  -86.8    & a  \\
  G331.512$-$0.066& 16:12:09.014 & -51:25:47.69& -108--132 & 0.11 &  -86.0    & a  \\
  G333.135$-$0.431& 16:21:02.821 & -50:35:12.01& -108--132 & 0.12 &  -50,-51.4& a,a\\ %
  {\em IRAS}16183$-$4958& 16:22:09.003 & -50:05:59.66& -94--117 & 0.12 &  -50.0    & d  \\
  G333.608$-$0.215& 16:22:11.043 & -50:05:56.52& -94--117 & 0.14 &  -51.6    & a  \\
  G338.280$+$0.542& 16:38:09.075 & -46:11:03.10& -107--133 & 0.10 &  -56.8    & a  \\
  G337.613$-$0.060& 16:38:09.543 & -47:04:59.89& -108--133 & 0.08 &  -42.0    & a  \\
  G337.705$-$0.053& 16:38:29.674 & -47:00:35.85& -108--133 & 0.09 &  -50.6    & a  \\
  G338.925$+$0.557& 16:40:33.490 & -45:41:37.00& -107--133 & 0.07 &  -61.3    & c  \\
  G339.622$-$0.121& 16:46:06.030 & -45:36:43.90& -107--133 & 0.07 &  -36.7    & c  \\
  G340.785$-$0.096& 16:50:14.838 & -44:42:26.71& -107--133 & 0.08 & -105.6,-102& a,c  \\
  G345.010$+$1.792& 16:56:47.580 & -41:14:25.73& -107--134 & 0.09 &  -21.5    & a  \\
  G343.929$+$0.125& 17:00:10.907 & -42:07:19.35& -107--133 & 0.08 &   13.6,11.9& a,b  \\
  G345.003$-$0.224& 17:05:11.205 & -41:29:06.96& -107--134 & 0.08 &  -26.2,-29.3& a,c  \\
  G345.698$-$0.090& 17:06:50.599 & -40:50:59.64& -106--133 & 0.08 &  -4.8     & a  \\
  G347.628$+$0.149& 17:11:50.888 & -39:09:29.00& -106--134 & 0.08 & -96.7     & a  \\
  G348.550$-$0.979& 17:19:20.418 & -39:03:51.65& -106--134 & 0.08 & -13.0,-12.1&a,c\\
  G351.417$+$0.645& 17:20:53.370 & -35:47:01.16& -106--135 & 0.09 & -10.0,-10.4&a,c\\
  {\em IRAS}17175$-$3544& 17:20:54.945 & -35:47:02.38& -106--135 & 0.09 & -10.0     & d  \\
  G351.581$-$0.353& 17:25:25.085 & -36:12:46.08& -106--135 & 0.07 & -93.8     & a  \\
  G351.775$-$0.536& 17:26:42.559 & -36:09:15.99& -106--135 & 0.09 & -8.7,4    & a,c  \\
  G353.410$-$0.360& 17:30:26.189 & -34:41:45.74& -105--134 & 0.08 & -20.7,-19 & a,c \\ 
  G354.724$+$0.300& 17:31:15.547 & -33:14:05.59& -105--135 & 0.08 &  90.2     & a  \\
  G355.344$+$0.147& 17:33:29.055 & -32:47:58.77& -104--135 & 0.08 &  18.0     & a  \\
  G359.138$+$0.031& 17:43:25.638 & -29:39:18.29& -104--136 & 0.10 &  -1.8     & a  \\
  G0.666$-$0.029& 17:47:18.640 & -28:22:54.60& -106--139 & 0.18 &  72.6     & a  \\
  G0.666$-$0.035& 17:47:20.143 & -28:23:06.18& -106--139 & 0.18 &  67.0     & a  \\  
  G10.473$+$0.027& 18:08:38.330 & -19:51:47.90& -100--140 & 0.09 &  64.7     & c  \\  
  G11.034$+$0.062& 18:09:39.770 & -19:21:23.60& -100--140 & 0.08 &  22.5     & c  \\  
  G11.904$-$0.141& 18:12:11.437 & -18:41:29.05& -100--140 & 0.09 &  42.8     & a  \\ %
  G10.623$-$0.383& 18:10:28.610 & -19:55:49.10& -100--140 & 0.09 &  -2.0,-1.5& b,e  \\
  G15.034$-$0.677& 18:20:24.811 & -16:11:34.14& -100--140 & 0.17 &  21.2     & a  \\ 
  W49N          & 19:10:12.006 & +09:06:11.25& -90--150  & 0.11 &   2.0     & d  \\ %
\hline
  \end{tabular}
  \label{tab:nondet4765}
  \end{minipage}
\end{table*}

A total of 55 star forming regions were searched for 4765-MHz OH
emission, the position, velocity range and noise level for each site
is given in Table~\ref{tab:nondet4765}.  Fourteen likely 4765-MHz OH
masers were detected towards nine different regions, two further sites
showed signs of broader thermal emission, or possibly blended weak
maser emission. These are listed in table \ref{tab:det4765} and
spectra of each source are shown in figure~\ref{fig:spec4765}. Ten of
the sources have not previously been detected as 4765-MHz OH masers,
the remaining previously reported in various papers, as indicated in
the table. All of the new detections are relatively weak and below the
sensitivity limit of most previous searches. We failed to detect
emission in six sites where 4765~MHz emission had previously been
reported (Mon~R2, {\em IRAS}12073-6233, {\em IRAS}16183-4958,
G338.925+0.557, {\em IRAS}17175-3544/NGC6334F, G10.623-0.383), a
number of these exhibit thermal emission observed with single dishes.
For {\em IRAS}16183-4958 and G10.623-0.383 the emission which has
previously been reported is broad and most likely thermal in nature
\cite{CMC95,GM83}. The thermal emission maybe resolved by our
interferometric observations, and then our detection limits would then
be considerably worse, as the sources would only contribute to the
shorter baselines. For those sources where the emission is strong
enough to plot intensity as a function of {\em uv} distance there is
no sign of this. However, none of these are thermal sources. The most
likely reason for the absence of emission from the other four sources,
which have clearly been detected as masers in previous observations,
is that the 4765-MHz maser is known to be highly variable, as reported
for example by Smits \etal\ \shortcite{SCH98}.

\begin{table*}
  \begin{minipage}{\textwidth}
  \caption{Sources with detected 4765-MHz OH emission. References : 
           *=new source, 
           a=Cohen, Masheder \& Caswell \protect\shortcite{CMC95};
           b=Gardner \& Ribes \protect\shortcite{GR71};
           c=Smits \protect\shortcite{S97};
           d=Zuckerman \& Palmer \protect\shortcite{ZP70}.}
  \begin{tabular}{lcccccrl}
    \hline
               & \multicolumn{2}{c}{\bf 4765-MHz Maser}    & {\bf Peak Flux} & 
    {\bf Velocity} & {\bf Width } &                   \\
  {\bf Source} & {\bf Right Ascension} & {\bf Declination} & {\bf Density}   &
    {\bf of Peak}  & {\bf of Peak}    &                   \\
  {\bf Name}   & {\bf (J2000)}         & {\bf (J2000)}     & {\bf (Jy)}      &
    {\bf (\kms)}   & {\bf (\kms)}   &  {\bf References} \\ \hline
  G240.316$+$0.071      & 07:44:51.982$\pm0.004$ & -24:07:42$\pm3$ &0.31,0.17&65.0,62.9& 
     0.4,0.4 &   * \\ 
  G240.311$+$0.074       & 07:44:52.00$\pm0.01$ & -24:07:22$\pm9$ &0.11&66.7& 
     0.4 &   * \\ 
  G294.511$-$1.621      & 11:35:32.232$\pm0.003$ & -63:14:43.12$\pm0.02$ &  2.02 & -12.0 & 
       0.3      &   c \\ 
  G309.921$+$0.479      & 13:50:41.77$\pm0.03$ & -61:35:09.9$\pm0.2$ &  0.17 & -60.9 & 
    2.1 &   * \\ 
  G328.304$+$0.436      & 15:54:03.94$\pm0.06$ & -53:11:32.2$\pm0.2$ & 0.12 & -91.8 & 
    0.4   &   * \\ 
  G328.307$+$0.430      & 15:54:06.38$\pm0.03$ & -53:11:41.0$\pm0.1$ & 0.16 & -90.5 & 
    0.9    &   * \\ 
  G328.808$+$0.633      & 15:55:48.33$\pm0.04$ & -52:43:06.5$\pm0.2$ &  0.17 & -44.6 & 
    0.3 &   * \\ 
  G328.809$+$0.633      & 15:55:48.42$\pm0.05$ & -52:43:06.2$\pm0.2$ &  0.14 & -43.4 & 
    0.5 &   * \\ 
  G333.135$-$0.431      & 16:21:02.80$\pm0.05$ & -50:35:10.1$\pm0.2$ &  0.14 & -51.2 & 
    0.8 &   * \\ 
  G333.135$-$0.431s     & 16:21:02.78$\pm0.04$ & -50:35:12.7$\pm0.2$ &  0.10 & -54.3 & 
    5.9 &   *,Thermal \\ 
  G353.410$-$0.360      & 17:30:26.182$\pm0.001$ & -34:41:46.15$\pm0.03$ & 1.78 & -20.8 &
       0.4      &   a,c \\ 
  Sgr~B2/G0.666$-$0.035 & 17:47:20.143 & -28:23:06.18 &  0.07 &  49.6 & 
     6.5 &   b, Thermal \\ 
  G11.904$-$0.141    & 18:12:11.437$\pm0.002$ & -18:41:29.0$\pm0.1$ & 1.0 &  41.8 & 
    0.4 &   * \\ 
  W49SW              & 19:10:10.9$\pm0.3$ &  +09:05:23$\pm5$ &  0.24,0.12 &   8.6,11.9 &  0.4,0.2     &   a,c \\ 
  W49N              & 19:10:13.51$\pm0.01$ &  +09:06:15$\pm2$ &  0.63 &  2.3 & 
       0.6      &   d,c \\ 
  W49NW             &  19:10:14.3$\pm0.2$ & +09:06:24$\pm3$  &  0.34 &   2.7 & 
       0.6      &   * \\ 
\hline
  \end{tabular}
  \label{tab:det4765}
  \end{minipage}
\end{table*}

\begin{figure*}
\dfig{g240.316+00.071}{g294.511-01.621}
\dfig{g309.921+00.479}{g328.307+00.430_2}
\dfig{g328.307+00.430}{g328.808+00.633}
\dfig{g333.135-00.431}{g333.135-00.431s}
\caption{Total intensity spectra of the 4765~MHz OH masers observed with 
  the ATCA}
\label{fig:spec4765}
\end{figure*}

\begin{figure*}
\dfig{g353.410-00.360}{g000.666-00.035}
\dfig{g011.904-00.141}{w49n}
\dfig{w49nw}{w49sw}
\contcaption{}
\end{figure*}

A total of 38 sites of 6035-MHz OH masers were searched for 4765-MHz
emission, resulting in ten detections (26\%). In comparison, of the 12
1720-OH masers searched, 2 were observed to harbour 4765-MHz emission
(17\%) and both of these are also sites of 6035-MHz OH masers. This
casts some doubt on the claimed predictive power of 1720/4765-MHz
association, suggesting the presence of other excited OH transitions
is at least as good a signpost for the presence of 4765-MHz OH masers,
although with few detections the statistics are not compelling.

\subsection{Comparison of different transitions} \label{sec:zeeman}
We have used our results to compare the emission we detect at 4765-MHz
with published observations of other OH transitions toward the same
site. This apparently simple task is complicated by the effect of
Zeeman splitting on the OH molecule, which shifts the observed
velocity by differing amounts for the various transitions.
Fortunately Caswell \& Vaile \shortcite{CV95} have determined the
magnetic field strength within the majority of the regions searched
from observations of the 6035-MHz masers. We have used this
information to calculate the velocity correction for each OH
transition of interest. We have calculated the velocity shift due to
Zeeman splitting for the various transitions using the method outlined
in Elitzur \shortcite{E92}, and using the Lande-g factors as measured
by Radford \shortcite{R61}. For the main-lines the Zeeman splitting is
straight forward with the different $\sigma$-components all
experiencing a velocity shift of essentially the same magnitude. For
satellite lines the situation is more complex with potentially
multiple Zeeman pairs being generated. From a theoretical standpoint,
it isn't clear which pair will dominate as this may depend upon the
pumping scheme, although observations tend to suggest that the pair
with the largest Einstein A-coefficients dominate, at least for the
1720-MHz transition \cite{GCRYF01}. Table~\ref{tab:velcor} lists the
amount by which RCP emission is shifted by the measured magnetic
fields in each of the regions which show 4765-MHz OH masers. We have
assumed that velocity shifts of 0.295, 0.177, 0.057 \&
0.028~km~s$^{-1}$~mG$^{-1}$ apply for the 1665-, 1667-, 1720- and
6035-MHz transitions respectively and that a positive $B$ field will
shift RCP emission to a more positive velocity \cite{CV95}. The
velocities in the table should therefore be subtracted from RCP
velocities and added to LCP velocities to correct for Zeeman
splitting. Where the alignment between different transitions is
discussed for the individual sources below, the velocities have been
corrected for the effects of Zeeman splitting (except where the
observation is of total intensity).

\begin{table*}
  \begin{minipage}{\textwidth}
  \caption{Velocity corrections for the effect of Zeeman splitting for
    RCP emission in the 1665-, 1667-, 1720- and 6035-MHz OH
    transitions. The magnetic field estimates listed are taken from
    Caswell \& Vaile \protect\shortcite{CV95}, except for W49N \& W49NW
    which are from Gaume \& Mutel \protect\shortcite{GM87}.}
  \begin{tabular}{lcrrrrr}
    \hline
                 &{\bf Magnetic} & {\bf 4765-MHz} &
    \multicolumn{4}{c}{\bf Velocity correction for OH transition (for RCP)} \\
  {\bf Source}     & {\bf Field}  & {\bf Velocity} & {\bf 1665-MHz} & 
    {\bf 1667-MHz} & {\bf 1720-MHz} & {\bf 6035-MHz} \\
  {\bf Name}       & {\bf (mG)}   & {\bf (\kms)}   & {\bf (\kms)}   & 
    {\bf (\kms)}   & {\bf (\kms)}   & {\bf (\kms)}   \\ \hline
  G240.316$+$0.071  & $<$ 0.5      &  +62.9,65.0  & $<$0.15     & $<$0.09     
    & $<$0.03     & $<$ 0.01    \\
  G294.511$-$1.621  & +1.1         &  -12.0  & 0.32        & 0.19        
    & 0.06        & 0.03        \\ 
  G309.921$+$0.479  & -2.5 -- +3.5 &  -60.9  & -0.74--1.03 & -0.44--0.62 
    & -0.14--0.20 & -0.07--0.10 \\ 
  G328.307$+$0.430  & 3.2          &  -90.5  & 0.94        & 0.57   
    & 0.18        & 0.09        \\ 
  G328.808$+$0.633  & -4.3         &  -44.6  & -1.27       & -0.76   
    & -0.24       & -0.12       \\ 
  G328.809$+$0.633  & -4.3         &  -43.4  & -1.27       & -0.76   
    & -0.24       & -0.12       \\ 
  G333.135$-$0.431  & -3.3         &  -51.2  & -0.97       & -0.58   
    & -0.19       & -0.09       \\ 
  G353.410$-$0.360  & -1.1         &  -20.8  & -0.32       & -0.19  
    & -0.06       & -0.03       \\ 
  G11.904$-$0.141   & $<$ 0.5      &  +41.8  & $<$0.15     & $<$0.09     
    & $<$0.03     & $<$ 0.01    \\
  W49SW           & -4.3         &  +8.6,11.9   &  -1.27      & -0.76
    & -0.24       & -0.12       \\ 
  W49N            & 3.1--7.5     &  +2.3   & 0.91--2.21  & 0.55--1.32
    & 0.18--0.42  & 0.09--0.21  \\ 
  W49NW           & 3.1--7.5     &  +2.7   & 0.91--2.21  & 0.55--1.32
    & 0.18--0.42  & 0.09--0.21  \\ 
  \hline
  \end{tabular}
  \label{tab:velcor}
  \end{minipage}
\end{table*}

We have used spectra in the literature to determine the strength of
the other OH maser transitions at the velocity of the 4765-MHz OH
maser emission. The results are listed in table~\ref{tab:fluxcom} and
demonstrate a number of general trends for the associations between
the 4765-MHz emission and other common OH maser transitions. With one
exception the 4765-MHz maser emission aligns in velocity with a peak
in the 6035-MHz spectrum and since we know that the two transitions
are spatially coincident to within 1 arcsecond then it seems likely
that the emission from the two transitions is co-spatial. The 1665-MHz
emission typically covers a wider velocity range than any of the other
transitions and so it is not surprising that it overlaps the velocity
range of the 4765-MHz emission. The 1665-MHz and 6035-MHz emission
which aligns with the 4765-MHz are usually not the strongest peak in
the spectrum. The 1665-MHz is often weaker than the 6035-MHz emission
at the same velocity, and 1667-MHz emission is usually weak or in some
cases entirely absent from the site. Caswell \shortcite{C96} found
81\% of main line OH maser sources have associated 6.7-GHz methanol
emission. We find for our sample of fourteen 4765-MHz OH masers only 61\%
have associated methanol masers. This may indicate a tendency for the
4765-MHz masers to be more common towards regions without 6.7-GHz
methanol masers, although the sample size is too small to reach
definitive conclusions. 

\begin{table*}
  \begin{minipage}{\textwidth}
  \caption{The flux density of 1665-, 1667-, 1720- and 6035-MHz OH
    transitions at the velocity of the 4765-MHz transition (after accounting
    for the shift due to Zeeman splitting). The reported
    flux densities are for RCP emission, except where (L)=LCP or (I)=Total
    intensity is indicated. Where no information is available for a particular
    transition than column has been left blank. References :
    a = Braz \etal\/ \protect\shortcite{BSGMG89};
    b = Caswell (pers. comm.);
    c = Caswell \protect\shortcite{C98};
    d = Caswell \protect\shortcite{C99};
    e = Caswell \& Haynes \protect\shortcite{CH83a};
    f = Caswell \& Haynes \protect\shortcite{CH83b};
    g = Caswell \& Haynes \protect\shortcite{CH87};
    h = Caswell \& Vaile \protect\shortcite{CV95};
    i = Caswell, Haynes \& Goss \protect\shortcite{CHG80};
    j = Gaume \& Mutel \protect\shortcite{GM87};
    k = Haynes \&  Caswell \protect\shortcite{HC77};
    l = MacLeod \protect\shortcite{M91}}
  \begin{tabular}{lcrrrrl}
    \hline
                 & {\bf 4765-MHz} &
    \multicolumn{4}{c}{\bf Flux Density at 4765-MHz Velocity} & \\
  {\bf Source}     & {\bf Velocity} & {\bf 1665-MHz} & {\bf 1667-MHz} & 
  {\bf 1720-MHz} & {\bf 6035-MHz} & \\
  {\bf Name}       & {\bf (\kms)}   & {\bf (Jy}   & {\bf (Jy)}   & 
  {\bf (Jy)}   & {\bf (Jy)}       & {\bf References}\\ \hline
  G240.316$+$0.071  &  +62.9  & 1.8     & $<$0.05 & $<$0.5 & 0.2    & 
  c,b,h,l \\
  G240.316$+$0.071  &  +65.0  & $<$0.05 & $<$0.05 & $<$0.5 & $<$0.1 & 
  c,b,h,l \\
  G294.511$-$1.621  &  -12.0  & 2.0     & 0.6(I)  &        & 4.6    & 
  a,b,c,h \\
  G309.921$+$0.479  &  -60.9  & 1.0     & 1.0     & $<$1.0 & 2.0    &
  g,h \\
  G328.307$+$0.430  &  -90.5  & 10.0(L) & $<$0.5  & $<$1.2 & 1.4    &
  h,i,k \\
  G328.808$+$0.633  &  -44.6  & 11.0(L) & 12.0(L) & $<$0.5 & 1.5    & 
  d,h,i \\
  G328.809$+$0.633  &  -43.4  & 6.0     & 1.5     & $<$0.5 & 5.0    & 
  b,d,h,i \\
  G333.135$-$0.431  &  -51.2  & 6.0     & 2.0     & $<$1.2 & 2.0    &
  i,h,k \\
  G353.410$-$0.360  &  -20.8  & $<$0.5  & $<$0.5  & $<$0.5 & 17.5   & 
  e,h,j \\
  G11.904$-$0.141   &  +41.8  & 0.5(L)  & 0.5     &        & 0.6    &
  f,h \\
  \hline
  \end{tabular}
  \label{tab:fluxcom}
  \end{minipage}
\end{table*}

\section{Notes on individual sources}

\subsection{Detections}

\subsubsection*{G240.316+0.071}
OH maser emission at this location was first detected in the 1665-MHz
transition by MacLeod \shortcite{M91}. Subsequent observations by
Caswell \shortcite{C98} detected weak 1667-MHz emission at the same
location, but offset in velocity from the 1665-MHz peak (Caswell,
pers.  comm.). The newly detected 4765-MHz maser has two spectral
features at velocities of 63.0 and 65.2~\kms\/, neither of which match
that of the strongest 6035-MHz emission at 63.6~\kms\/ \cite{CV95}.
The secondary peak of the 6035-MHz emission matches the weaker
4765-MHz feature at 62.8~\kms. There is no 6.7-GHz methanol maser
emission towards this site.

\subsubsection*{G240.311+0.074}
This newly detected maser source is offset to the south by
approximately 20\arcsec\/ from G240.316+0.071. High resolution
observations by Caswell at 1665- and 6035-MHz \shortcite{C97,C98} have
not reported any emission towards this location. A search of SIMBAD
reveals no other astrophysical objects within 10\arcsec\/ of this site.

\subsubsection*{G294.511-1.621}
The ground-state OH maser emission at this location was detected by
Braz \etal\/ \shortcite{BSGMG89} towards an \water\/ maser source. The
4765-MHz OH maser emission was discovered by Smits \shortcite{S97} and is
the strongest detected in this survey with a peak flux density of
2~Jy. The velocity and position of the 4765-MHz emission matches that
of the 6035-MHz emission at -12~\kms\/ \cite{CV95}. A minor peak of
the 1665-MHz aligns with the 4765-MHz maser and there is probably
1667-MHz emission at the same velocity, but the ATCA observation is of
total intensity and so cannot be corrected for Zeeman splitting
(Caswell, pers. comm.). 

\subsubsection*{G309.921+0.479}
The 4765-MHz OH emission in this source is relatively broad (2.1~\kms)
and may be quasi-thermal emission or a blend of a number of weak maser
features. The secondary peak (-59.8~\kms) matches the peak of the
6035-MHz maser, the broad emission is echoed in the weaker broad
emission in the 6035-MHz spectrum \cite{CV95}. The magnetic field in
this region is confused and so determining the alignment for the 1665-
and 1667-MHz transitions is less certain, it appears that in both
cases the 4765-MHz emission aligns with lesser peaks \cite{CH87}.

\subsubsection*{G328.304+0.436}

This newly detected maser source is offset to the south-west by
approximately 24\arcsec\/ from G328.307+0.430. As for G240.311+0.074,
Caswell does not report any emission at 1665- or 6035-MHz
\shortcite{C97,C98} towards this location, and a SIMBAD search finds
nothing within 10\arcsec.

\subsubsection*{G328.307+0.430}
This new 4765-MHz detection has two features at -90.5 and -92.0
\kms\/, it was not detected in previous surveys, the most sensitive of
which was Cohen \etal\/ \shortcite{CMC95}. The 1667-MHz emission
towards this source is weak and unlike the majority of the other
regions there is no 6.7-GHz methanol maser emission at this site. The
stronger 4765-MHz emission matches the velocity of a 10~Jy LCP peak at
1665-MHz, although there is no Zeeman pair \cite{CHG80}. The 6035-MHz
emission is also aligned in velocity with the strongest 4765-MHz
maser.

\subsubsection*{G328.808+0.633/G328.809+0.633}
Caswell \shortcite{C97} reports two nearby sources each of which
exhibit both 6035~MHz OH and 6.7-GHz methanol masers. To within the
accuracy of our resolution we find 4765-MHz OH maser emission to also
be present at both sites. G328.809+0.633 is offset from the \ionhy\/
region and also exhibits 1720-MHz OH emission. The velocity of the
1720-MHz OH maser in the total intensity observation of Caswell
\shortcite{C99} is offset by +0.2~\kms\/ from the 4765-MHz peak. Full
polarization observations reveal that after correction for Zeeman
splitting the 1720-MHz emission does not align with the 4765-MHz maser
peak (Caswell pers. comm.). There appears to be peaks in the 6035-MHz
spectrum which correspond to the two 4765-MHz peaks \cite{CV95} and
the strongest 1665-/1667-MHz emission covers the same velocity range
\cite{CHG80}.

\subsubsection*{G333.135-0.431/G333.135-0.431s}
There are two sites of 6035-MHz OH maser emission in this region.
G333.135-0.431 contains the stronger emission and doesn't have an
associated 6.7-GHz methanol maser, while G333.135-0.431s contains
weaker 6035-MHz emission covering a larger velocity range and with
weak 6.7-GHz methanol masers \cite{C97}. The strong narrow 4765-MHz
maser lies towards the first of these sites and matches the velocity
of secondary peaks in all of the 1665-, 1667 and 6035-MHz transitions
\cite{CHG80,CV95}. The 4765-MHz emission towards G333.135-0.431s
appears to be thermal and matches the velocity of the strongest
1665-/1667-MHz absorption, broad weaker 6035-MHz emission and broad
weak 6.7 GHz methanol emission \cite{CHG80,C97}.

\subsubsection*{G353.410-0.360}

This was the strongest source detected by Cohen \etal
\shortcite{CMC95}, and has not varied more than 6\%. The 4765-MHz peak
matches the velocity and position of the peak of the complex 6035-MHz
spectrum, but it isn't clear whether or not the 1720-MHz emission
matches the 4765-MHz velocity. The observations of Gaume \& Mutel
\shortcite{GM87} have coarse spectral resolution, while the total
intensity observations of Caswell \shortcite{C99} show a significant
velocity offset, suggesting that the transitions probably do not
align. The 1665- and 1667-MHz velocity ranges do not overlap with the
4765-MHz emission \cite{CH83a}.

\subsubsection*{Sgr~B2/G0.666-0.035}

4765-MHz maser emission was first observed towards Sgr~B2 by Gardner
\& Ribes \shortcite{GR71} at 60.8 and 61.4~\kms\/ using the Parkes
telescope. It was not seen by Effelsburg or VLA observations in 1983
and 1984 \cite{GM83,GWP87}. We find no narrow emission, but a weak and
broad feature at a lower velocity which matches the broad-band
observations of Gardner \& Ribes \shortcite{GR71}. This does not match
the velocity of the 6035-MHz or 1720-MHz maser emission
\cite{CV95,GM87}.

\subsubsection*{G11.904-0.141}
The 4765-MHz maser emission from this new detection peaks at 41.8~\kms\/
and aligns in velocity with the weaker of the two 6035-MHz maser peaks
\cite{CV95}. It also matches the velocity of the strongest LCP
features in the 1665- and 1667-MHz spectra \cite{CH83b,FC89}.

\subsubsection*{W49N and W49SW}
Previous single dish observations have not been able to discriminate
between the 4765-MHz OH emission at 2.2~\kms\/ from the northern
portion of this region with the emission at 8.2~\kms\/ from the south
west portion, 1.2\arcmin\/ away. Two features can be found in W49N, the
stronger peaking at 2.1~\kms\/, the other peaking at 2.6~\kms\/. The
two sources are offset by 12\arcsec\/ to the East. 
Smits \shortcite{S97} has summarised the previous observations of the
4765-MHz OH maser emission in this source, and we detect all the
spectral features observed by him and confirm a velocity of 8.6~\kms\/
for the emission towards W49SW. We also detect a previously
unreported maser towards the same region at a velocity of 11.9~\kms.
No feature was observed corresponding to the report of Smits
\shortcite{S97} of a weak broad emission around 3.6~\kms, but the
intensity of all the observed features appears to have declined in the
intervening period. The OH maser emission in this region is very
complex and we have not attempted to determine the alignment of the
4765-MHz emission with other transitions as the potential for
confusion is too great. This is exacerbated by the fact that for these
northern sources our positional errors in declination are several
arcseconds. However, the position Caswell \& Vaile \shortcite{CV95}
report for the strongest 6035-MHz emission matches the 4765-MHz maser
W49SW.



\subsection{Non detections}

\subsubsection*{Mon~R2}


Mon~R2 is one of the best studied 4765-MHz masers and is known to be
highly variable and also show polarized emission \cite{SCH98}.
Emission from the 4765-MHz transition towards this source was first
reported by Gardner \& Martin-Pintado \shortcite{GM83}. Cohen
\etal\/ \shortcite{CMW91} failed to detect any emission, but it was
later detected by Cohen \etal\/ \shortcite{CMC95}. Smits \etal\/
monitored it closely from late 1994 through to mid-1997 and it reached
a peak flux density of 80~Jy in 1997, at times doubling in intensity
every 19 days \cite{SCH98}. When observed by Szymczak \etal\/
\shortcite{SKH00} in 1998 it had declined by an order of magnitude from its
peak, to a flux density 6~Jy and it subsequently declined further to
below the detection limits for the Hartebeesthoek radio telescope
(Smits, pers. comm.). We also find no emission in this survey with
an upper limit of 80~mJy, 3 orders of magnitude lower than its strongest
recorded flux density.

\subsubsection*{{\em IRAS}12073-6233}

This source was discovered by Cohen \etal\/ \shortcite{CMC95} and
subsequently monitoring by Smits \shortcite{S97} observed it to reach a
maximum flux density of 1.2~Jy in 1994.8, which afterwards fell at about
0.5 Jy per year. Our non-detection of this source is consistent with
this decline continuing.

\subsubsection*{{\em IRAS}16183-4958}
This weak detection in the survey of Cohen \etal\/ \shortcite{CMC95},
was not detected in our observations. Had it remained constant it
would have been well above our detection limit and so must have
decreased in intensity by at least 50\% in the intervening period.

\subsubsection*{G338.925+0.557}
This source was detected in a search of 1720-MHz OH maser sources by
MacLeod \shortcite{M97}. MacLeod reports a peak flux density of
1.5~Jy, while we fail to detect any emission stronger than 70~mJy,
indicating that this source decreased in intensity by more than a
factor of 20 in the four year period between the two sets of
observations.

\subsubsection*{{\em IRAS}17175-3544/NGC6334F}
This weak detection in the survey of Cohen \etal\/ \shortcite{CMC95},
was not detected in our observations. If it had remained constant it
would have been well above our detection limit and so must have
decreased in intensity by at least 67\% in the intervening period.

\section{Discussion} 

The association between 1720- and 4765-MHz OH masers has only been
studied in detail towards W3(OH). Of the two 4765-MHz masers we
detected which may be closely associated with 1720-MHz OH emission,
for one the velocity (after correction for Zeeman splitting) doesn't
match and for the other it may align, but further investigation is
required. VLBI observations of the 1720-MHz OH masers by Masheder
\etal\/ \shortcite{MFGMCB94} found that although the two transitions
came from the same spatial location (within the relative
uncertainties), the velocities did not align. Intriguingly, more
recent MERLIN observations by Gray \etal\/ \shortcite{GCRYF01} found
both spatial and velocity alignment between the two transitions. One
possible explanation for this is that the 4765-MHz OH masers can be
highly variable and it is not unknown for masers features to disappear
and new features appear at nearby, but different velocities
(e.g. W49N). This highlights the major pitfall in attempting to test
theoretical models with multi-transition observations; variability
over time and the effects of observations at different spatial and
spectral resolutions all add uncertainty to comparisons of the
different transitions. Our observations suggest that 4765-MHz OH
masers are frequently not accompanied by 1720-MHz OH masers, but
further high resolution, full polarization observations of the
1720-MHz OH masers in southern star forming regions are required to
confirm this.

It is clear from literature that current theoretical OH maser models are
complicated by unknown, or poorly known excitation rates for the
pumping paths. The models of Gray \etal\/ \shortcite{GFD92} produce
parameter ranges where both 1720- and 4765-MHz transitions are
inverted with local temperatures of 125K, high densities and large
velocity gradients. These are the only conditions under which
4765-MHz masers are significantly excited and the 6035-MHz line is
strongly suppressed under these conditions. This is contrary to our
observations in which 4765- and 6035-MHz masers are commonly present
together. 

In contrast the models of Pavlakis \& Kylafis \shortcite{PK96b}, which
use different collision rates find 4765-MHz masers are only excited
with {\em low} velocity gradients, while 1720-MHz emission is only
exited with {\em high} velocity gradients. Pavlakis \& Kylafis
\shortcite{PK96b} find that the 4765-MHz masers only appear at higher
densities than the ground state main lines and there is essentially no
overlapping density range. This contrasts with our observations that
the 4765-MHz OH masers are typically associated with the 1665-MHz
transition. Modelling of the 6035-MHz OH masers is covered in a
separate paper \cite{PK00} and the issue of the association between
4765- and 6035-MHz masers is not covered. Through comparison of the
plots from the two papers we can deduce that the two transitions are
both inverted for the narrow molecular hydrogen density range of $4
\times 10^6 - 10^7$cm$^{-3}$, with 150~K gas, 200~K dust temperatures and large
optical depth (Pavlakis \& Kylafis 1996b - Fig. 6d ; 2000 - Fig. 9d).
Once again, for these physical conditions the ground state main lines
are not inverted. Pavlakis \& Kylafis have many more free parameters
in their modelling, complicating interpretation. It is also possible that
there are, unpublished parameter combinations that excite 4765-, with
1665-, 1720- and/or 6035-MHz.

Pavlakis and Kylafis \shortcite{PK96b} find a number of different parameter
combinations which produce 1720-MHz OH masers, but the strongest
emission occurs in the absence of far infrared emission, at high
densities and large velocity gradients. Under these conditions they
also predict that the 4750-MHz transition should produce bright masers.
Our search failed to detect any strong 4750-MHz masers associated with
1720-MHz OH masers, which suggests that the one of the other parameter
combinations must be responsible for the 1720-MHz OH masers.


\section{Conclusions} 

We report the discovery of ten new $^2\Pi_{1/2}$ F=1$\rightarrow$0
4765-MHz OH maser emission sites,
more than doubling the number reported in the southern hemisphere. We
have found that the frequently purported association between 1720-MHz
OH masers and 4765-MHz emission is not as strong as the correspondence
between 6035- and 4765-MHz masers. The 4765-MHz OH masers are
relatively weak and uncommon compared to the 1665-, 1667- and 6035-MHz
transitions and so, except for attempting to explain the
1720-/4765-MHz association, the models of Pavlakis \& Kylafis
\shortcite{PK96a,PK96b,PK00} and Gray \etal\/
\shortcite{GFD92,GCRYF01} largely ignore this transition. Examination
of published spectra of the 1665-, 1667-, 1720- and 6035-MHz OH
transitions shows that after correcting for the effects of Zeeman
splitting the 4765-MHz OH masers are characterised by a close
association with the 6035-MHz transition, modest 1665-MHz emission and
weak or absent 1667-MHz emission. None of the current models of OH
maser emission produces results which are qualitatively consistent
with these findings. We suggest that the best indicator of likely
4765-MHz OH masers is the presence of other excited state transitions,
such as the 6035-MHz, but that these regions also frequently possess
the correct physical conditions to produce satellite-line masers from
the ground-state. Further theoretical modelling is required to
determine ranges of physical parameters which can produce OH maser
emission with the same characteristic as we observe. In addition, we
failed to find any sites where 1720-MHz and 4750-MHz OH emission are
produced simultaneously despite this being a well populated condition in
theoretical models for certain parameter ranges.

Determining the relative intensity of emission from different OH
transitions by comparison with published observations has several
major sources of uncertainty, in particular temporal variability and
instrumental effects. To overcome this it would be highly desirable
to be able to make simultaneous, or quasi-simultaneous full
polarization observations of a large number of OH transitions, at high
spatial and spectral resolution. Such observations are possible, for
example, with the ATCA. The 19 lowest frequency OH transitions fall
within the frequency coverage and it is possible to observe many, or all
of these within a single observing session. This type of observation
would clearly provide much stronger observational constraints and
tests for theoretical modelling than are currently available.

A number of the newly discovered 4765-MHz OH masers (G240.331+0.074
\&G328.304+0.436) are not associated with previously detected sites of
ground-state OH masers or other common sign-posts of star formation,
such as {\em IRAS} sources. These masers are very weak, but
further investigation is warranted to confirm their existence and to
try and determine whether they are associated with high-mass star
formation, or some other astrophysical phenomena.

\section{Acknowledgments}

We would like to thank J.L. Caswell for providing unpublished Parkes
and ATCA spectra for a number of sources and communication of
unpublished results. The Australia Telescope Compact Array is part of
the Australia Telescope, funded by the Commonwealth of Australia for
operation as a National Facility, and managed by CSIRO. This research
has made use of the SIMBAD database operated at CDS, Strasbourg,
France. This research has made use of NASA's Astrophysics Data System
Abstract Service.

\end{document}


\title[Erratum: Search for 4765-MHz OH masers] {Erratum: A search
for 4765-MHz OH masers in Southern Star Forming Regions}

\author[Dodson, Voronkov \& Ellingsen]{R. Dodson $^1$, M.A. Voronkov $^{2,3}$,
  S.P. Ellingsen $^4$ \\
$^1$ ISAS, Yoshinodai 3-1-1 , Sagamihara, Kanagawa
229-8510, Japan \\
$^2$ Australia Telescope National Facility, CSIRO,Locked Bag 194, Narrabri, NSW
 2390, Australia\\
$^3$ Astro Space Centre, Profsouznaya st. 84/32, 117997 Moscow, Russia;\\
$^4$ School of Mathematics and Physics, University of Tasmania, 
     GPO Box 252-21, Hobart, Tasmania 7001, Australia}

\maketitle

\begin{abstract}
  
  Dodson \& Ellingsen (2002) included several observations with
  significant pointing errors, invalidating the upper limits found in
  these directions. These have now been reobserved or recalculated. A
  new table of upper limits has been generated, and two more masers
  that would have been seen have been found.

\end{abstract}

\begin{keywords}
H{\sc ii} regions --- ISM: molecules --- masers --- radio lines : ISM
\end{keywords}

\section{Observations}

Our followup to the original observations \cite{DE02}, performed on 15
December 2003, with the ATCA in configuration 6A, had only 10
baselines so the detection limits are a little higher than the
original data set and are listed in Table 1.
As before the detection peak heights, velocities and widths are found
from a Gaussian fit to the unsmoothed data. The detections are listed
in Table 2.


\subsubsection*{G339.622-0.121}


The 4765-MHz maser emission from this new detection peaks at
-37.0~\kms . This site has 1720-, 1665- and 6035-MHz OH masers \cite{C99}.
There is a nearby 1612-MHz emission site which also displays methanol
masing. In this region there is no detectable 4765-MHz emission.
The flux from this source is below the limit of Smits \shortcite{S03}.

\subsubsection*{G347.628+0.148}


This source was discovered by Smits \shortcite{S03}, with a slightly stronger
flux density (a peak of 5 Jy). As the source is very narrow (0.3 km/s
FWHM) our observations may underestimate the strength, but the
measured emission was falling during the Hartsebethock observations
and is consistent. This source has a peak velocity of -96.3 \kms\ and
a secondary peak of 0.3~Jy at -95.1 \kms\ from the same position.
This source also has emission from 1612-, 1665-, 1720- and 6035-MHz, and
6.7-GHz methanol. It is an \UCHII\ region.

\begin{table*}
  \begin{minipage}{\textwidth}
  \caption{Sources searched for 4765-MHz OH maser emission. The 5-$\sigma$ 
    limit column is for emission in a vector averaged spectrum at centred on 
    the position, and covering the velocity range listed. References :
    a = Caswell \protect\shortcite{C97} ; 
    c = Caswell \protect\shortcite{C99} ;
    d = Cohen \etal\/ \protect\shortcite{CMC95} ; }
  \begin{tabular}{lcccrcl}
    \hline
               & \multicolumn{2}{c}{\bf Reference position} & {\bf Velocity} & 
    {\bf 5-$\sigma$} & {\bf Central}  & 
                     \\
  {\bf Source} & {\bf Right Ascension} & {\bf Declination} & {\bf Range}    &
    {\bf Limit}           & {\bf Velocity} & 
                     \\
  {\bf Name}   & {\bf (J2000)}         & {\bf (J2000)}     & {\bf (\kms)}   &
    {\bf (Jy)}            & {\bf (\kms)}   & 
    {\bf References} \\ \hline
\multicolumn{7}{c}{reobserved}\\
  G339.622$-$0.121& 16:46:06.030 & -45:36:43.90& -130--95 & 0.11 & -36.7    & c  \\
  G347.628$+$0.149& 17:11:50.888 & -39:09:29.00& -130--95 & 0.12 & -96.7     & a  \\
  G348.550$-$0.979& 17:19:20.418 & -39:03:51.65& -130--95 & 0.12 & -13.0,-12.1&a,c\\
  G351.581$-$0.353& 17:25:25.085 & -36:12:46.08& -130--95 & 0.15 & -93.8     & a  \\
\multicolumn{7}{c}{corrected}\\
  G351.417$+$0.645& 17:20:53.370 & -35:47:01.16& -106--135 & 0.11 & -10.0,-10.4&a,c\\
  {\em IRAS}17175$-$3544& 17:20:54.945 & -35:47:02.38& -106--135 & 0.11 & -10.0     & d  \\
  G351.775$-$0.536& 17:26:42.559 & -36:09:15.99& -106--135 & 0.15 & -8.7,4    & a,c  \\
\hline
  \end{tabular}
  \label{tab:nondet4765}
  \end{minipage}
\end{table*}

\begin{table*}
  \begin{minipage}{\textwidth}
  \caption{Sources with detected 4765-MHz OH emission. References : 
           *=new source, 
           e=Smits \protect\shortcite{S03};}
  \begin{tabular}{lcccccrl}
    \hline
               & \multicolumn{2}{c}{\bf 4765-MHz Maser}    & {\bf Peak Flux} & 
    {\bf Velocity} & {\bf Width } &                   \\
  {\bf Source} & {\bf Right Ascension} & {\bf Declination} & {\bf Density}   &
    {\bf of Peak}  & {\bf of Peak}    &                   \\
  {\bf Name}   & {\bf (J2000)}         & {\bf (J2000)}     & {\bf (Jy)}      &
    {\bf (\kms)}   & {\bf (\kms)}   &  {\bf References} \\ \hline
  G339.622$-$0.121&16:46:06.03$\pm0.04$ & -45:36:43.3$\pm0.2$ & 0.56 & -37.0 & 
    0.5 &   * \\ 
  G347.628$+$0.149&17:11:51.01$\pm0.02$ & -39:09:29.0$\pm0.5$ & 3.65 & -96.3 & 
    0.3 &   e \\ 
\hline
  \end{tabular}
  \label{tab:det4765}
  \end{minipage}
\end{table*}

\begin{figure*}
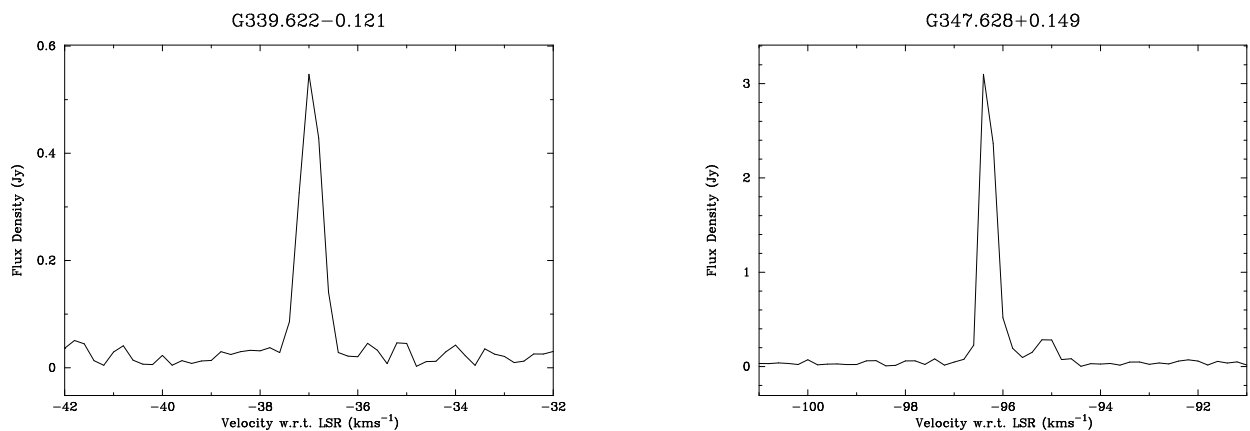

\dfig{g339.622-0.121}{g347.628+00.149}
\caption{Total intensity spectra of the 4765~MHz OH masers reobserved with 
  the ATCA}
\end{figure*}

\section{conclusions}

Both sources found also display 1720- and 6035-MHz OH emission, so they do
not significantly alter our previous conclusions about the association
of 4765-MHz OH emission with these two transitions.

\section{Acknowledgments}
We would like to thank Derck Smits for drawing our attention to the
error in the original paper.